\documentclass[10pt, conference, compsocconf]{IEEEtran}

\usepackage{amsmath}
\usepackage{amssymb}
\usepackage{graphicx}
\usepackage{subfig}
\usepackage{multirow}

\begin{document}

\title{RLWS: A Reinforcement Learning based GPU Warp Scheduler}
\author{
\IEEEauthorblockN{Jayvant Anantpur}
\IEEEauthorblockA{Mentor Graphics, Bangalore, India\\
jayvant.anantpur@gmail.com}
\and
\IEEEauthorblockN{Nagendra Gulur Dwarakanath}
\IEEEauthorblockA{Texas Instruments,
Dallas, USA \\
nagendra.gd@gmail.com}
\and
\IEEEauthorblockN{Shivaram Kalyanakrishnan}
\IEEEauthorblockA{Dept. of CSA,
IIT Bombay,
Mumbai, India \\
shivaram@cse.iitb.ac.in}
\and
\IEEEauthorblockN{Shalabh Bhatnagar}
\IEEEauthorblockA{Dept. of CSA,
IISc,
Bangalore, India \\
shalabh@csa.iisc.ernet.in}
\and
\IEEEauthorblockN{R. Govindarajan}
\IEEEauthorblockA{SERC,
IISc,
Bangalore, India \\
govind@serc.iisc.ernet.in}
}

\maketitle

\begin{abstract}

The Streaming Multiprocessors (SMs) of a Graphics Processing Unit (GPU) 
execute instructions from a group of consecutive threads, called warps. At each cycle, an SM schedules a warp from a group of active warps and can context switch among the active warps to hide various stalls. 
Hence the performance of warp scheduler is critical to the performance of GPU. Several heuristic warp scheduling algorithms have been proposed which 
work well only for the
situations they are designed for. GPU workloads are becoming very diverse in nature
and hence one heuristic may not work for all cases. To work well over a diverse
range of workloads, which might exhibit hitherto unseen characteristics, a warp scheduling
algorithm must be able to adapt on-line.

We propose a Reinforcement Learning based Warp Scheduler (RLWS) which learns to
schedule warps based on the current state of the core and the long-term benefits of
scheduling actions, adapting not only to different types of workloads, but also to different execution phases in each workload.  As the design space involving the state variables and the 
parameters (such as learning and exploration rates, reward and penalty values) used by RLWS is large, we use Genetic Algorithm to identify the useful subset of state variables and parameter values.
We evaluated the
proposed RLWS using the GPGPU-SIM simulator on a large
number of workloads from the Rodinia, Parboil, CUDA-SDK and GPGPU-SIM benchmark
suites and compared with other state-of-the-art warp scheduling methods. Our RL based implementation achieved either the best or very close to the best performance in 80\% of kernels with an average speedup of 1.06x over
the Loose Round Robin strategy and 1.07x over the Two-Level strategy. 

\end{abstract}

\begin{IEEEkeywords}
GPU; Warp Scheduling; Divergence;

\end{IEEEkeywords}

\section{Introduction}
\label{sec:Introduction}
Graphics Processing Units (GPUs) have proved to be highly effective and energy efficient for accelerating regular data-parallel
applications. With improvements in technology and smaller transistor sizes, latest GPUs have 
thousands of processing cores, organized as multiple Streaming Multiprocessors (SMs)\footnote{We use the terminology of NVIDIA GPUs and Compute Unified Device Architecture (CUDA) throughout the paper. However the ideas proposed in the paper are equally applicable to ATI Radeon GPUs and OpenCL.}, each comprising of a set of SIMD cores. The SIMD cores together execute a group of consecutive threads, called warp. To hide stalls due to data dependency or  memory operations,  an SM can context switch on each cycle from among a set of active warps.

The processing units in a GPU issue instructions in-order, 
have very small cache capacity per thread, simple or no branch prediction ability, etc. These factors make effective warp scheduling critical to the performance of the GPU. Existing
warp schedulers address only a subset of these factors. The two-level (TL) \cite{TwoLevel} and Greedy Then Old (GTO) warp schedulers, for example, address the problem of hiding long latencies by staggered execution of warps.  
In addition to long latencies, some other factors affecting warp schedulers are branch divergence \cite{DynWarpSubdivision},  warp divergence \cite{WarpLevelDivergence}, memory and cache contention \cite{NMNL}, \cite{CCWS}, etc. Researchers have proposed a number of heuristics for these factors \cite{Jog}, \cite{OWL}, \cite{ThreadFrontier}, \cite{DynWarpSubdivision}, \cite{DualPath}, \cite{NMNL}, \cite{WarpLevelDivergence}, 
\cite{CAWS}, \cite{CAWA}, \cite{DAWS},  \cite{CCWS}, \cite{TwoLevel}, \cite{ThreadBlockCompaction}, \cite{DynWarpForm}, 
\cite{Mascar}, \cite{PriorityCache}.
But such heuristics improve performance only on a subset of applications \cite{iPAWS}, \cite{PhaseAware}. They cannot effectively handle different kernel types and also different execution phases in a kernel. With GPU workloads becoming diverse in nature, a warp scheduling algorithm must be robust, applicable to different phases of execution and workloads, and be able to adapt on-line.

On the other hand, tremendous amount of similarity is exhibited among the execution of warps of a thread block and among thread blocks of a kernel. This similarity is in terms of dependency across instructions, memory and synchronization latencies of instructions, use of special functional units (SFUs) and control flow in a warp. Since the warps are executed one group (of active warps) at a time, can a warp scheduler be designed to learn from the scheduling decisions taken in the current set of warps and deploy the same in future warps? In an attempt to address this, in this paper, we propose a warp scheduler which uses Reinforcement Learning (RL) \cite{Sutton} to explore different scheduling actions, learn from
them and adapt to changing application requirements. RL based schedulers can be constructed as autonomous agents to learn optimal actions by observing 
their environments. The feedback given by the environment enables RL based schedulers to adapt to the situations and change their
actions. 

The environment of an RL based warp scheduler consists of many factors, such as current state of resident warps, caches, execution pipelines, etc. Each of these factors can take a number of values, drastically increasing the number of different states the environment can be in. In addition to effectively characterizing the environment, for an RL system to work well, parameters such as exploration rate, learning rate, rewards, etc., need to be assigned appropriate values. These make the design space of the parameters used by RLWS very large. Hence we used genetic algorithm to systematically search the exponential design space and obtain the near-optimal RL design.

In this paper, we discuss in detail the design of a Reinforcement Learning based Warp Scheduler (RLWS) as a generic warp scheduler as opposed to heuristic based warp schedulers. 
We explain various techniques used to reduce hardware overheads of RLWS.

A detailed evaluation on a large and diverge set of 59 kernels from Rodinia \cite{Rodinia}, Parboil \cite{Parboil}, GPGPU-SIM \cite{GPGPUSIM} and CUDA SDK \cite{SDK} benchmark suites showed encouraging results. For about 80\% (47 out of 59) kernels, RLWS is either the best or second best compared to other competitive warp schedulers, proving its robustness. RLWS achieves a geometric mean performance improvement of 1.06x over LRR (Loose Round Robin), 1.07x over TL, 1.02x over iPAWS \cite{iPAWS} and matches the performance of GTO. RLWS performs better than LRR on GTO friendly kernels and better than GTO on LRR friendly kernels.

With increasing use of GPUs in accelerating a wide range of applications, a generic and robust warp scheduler design like RLWS opens up new opportunities for optimizations. Since RLWS can adapt to different situations, scheduler designers can focus on identifying key state variables and actions to improve the agent-environment interaction, instead of designing a fixed policy. 
To the best of our knowledge, our work is the first to use the concept of RL to design a warp scheduler for GPUs.

\section{Background}
\label{sec:Background}
In this section we give a brief background on GPUs, Reinforcement Learning and Genetic Algorithm.

\subsection{Graphics Processing Unit}
Graphics Processing Units (GPUs) are being used in general purpose computing to accelerate both regular and irregular applications.  
Two most commonly used languages for programming GPUs are CUDA \cite{CUDA} and OpenCL \cite{Opencl} which use Single
Instruction Multiple Threads (SIMT) computation model \cite{CUDA}. In this model a large number of threads are run in parallel on 
Single Instruction Multiple Data (SIMD) cores using hardware multi-threading. A group of SIMD cores constitute a Streaming Multiprocessor (SM) in NVIDIA GPU and a GPU typically consists of multiple SMs. For example, the Kepler GK110 GPU \cite{Kepler} consists of 15 SMs where each SM contains 192 CUDA cores with each core 
containing a fully pipelined integer arithmetic logic unit and floating point 
unit. In addition, each SM also contains 32 Load/Store Units, 32 Special Function Units, 64 Double Precision units, a register file of 64K 4 byte registers, L1-D cache and software managed shared memory.

A typical usage of GPU involves defining functions in languages like CUDA, OpenCL, etc., and invoking those functions from a thread
running on a CPU. These functions are known as \emph{kernels}. A kernel is invoked with an execution configuration called grid, which 
describes the hierarchical structure of threads that will execute the kernel code \cite{CUDA}. A grid is a 1-D, 2-D or 3-D structure of thread blocks where
a thread block (TB) is a 1-D, 2-D or 3-D structure of threads.
The threads of a TB can communicate and cooperate with
each other using shared memory and barrier synchronization. TBs run independent of each other. 

Each SM has a limited number of resources such as registers, shared memory,  etc.,  and hence the number of thread
blocks that can be allocated to an SM depends on the resources required by each thread and TB. The resource allocation and 
deallocation is done at the TB level. A global work distribution engine (\emph{Thread Block Scheduler}) in the GPU assigns each TB to an SM.
On a kernel invocation, the TB scheduler will assign as many TBs to SMs as allowed by
resource constraints and then assign the remaining TBs one at a time as and when a previously assigned TB finishes. 

The threads of a TB are further partitioned into groups of consecutive threads, called \emph{warps} in the CUDA terminology. The size
of a warp is 32 threads. Each SM contains one or more warp schedulers, each of which schedules one ready warp and issues the next 
instruction to the execution pipeline. For example, Fermi GPU \cite{Fermi} has two warp schedulers and each scheduler can issue one warp every cycle. Due to limited number of Load/Store and Special Function Units(SFU), each SM can issue only one memory or one SFU instruction per cycle, but it can issue two integer or
floating point instructions --- one by each warp scheduler --- which are executed
on the 32 CUDA cores. We refer to these three instructions as MEM, SFU and SP
instructions, and the corresponding pipelines as MEM, SFU and SP pipelines.

Instructions of warps are issued \emph{in order}. If the current instruction of a warp can not be issued for any reason, then the warp is not ready for scheduling. With fewer ready warps, the ability of a warp scheduler to overlap execution of instructions from different warps and hide long execution latencies decreases.
A cycle in which the warp scheduler can not issue any warp is called a stall cycle. Stalls can be caused because of many factors such as unavailability of data, unavailability of execution pipeline resources, empty instruction buffers, etc. Some of the reasons behind these are cache misses, long execution latencies, barrier statements, allocation and deallocation of threads at TB level, etc. For example, when a warp issues a load memory instruction which misses in L1-D and L2 caches, it can take hundreds of cycles before the load completes. An efficient warp scheduler can hide much of the stalls experienced by threads through effective context switching across other ready warps. Thus the performance of the warp schedulers is crucial for the performance of the GPU. 

\subsection{Reinforcement Learning}
Reinforcement Learning is a machine learning technique in which, autonomous agents learn through interactions with their environment \cite{Sutton}.
Every time step the agent receives as input a representation of the environment state. Based on the state, it takes an action which may change the environment state. As a consequence of the action, it receives a numerical reward from the environment. The goal of the agent is to learn a policy --- a mapping from states to actions --- that maximizes the total rewards it receives, over its lifetime. The agent is not told which actions to take, but learns, by trying them, the most rewarding actions.

There are multiple challenging aspects of RL. The RL agent needs to learn which
actions are most rewarding and needs to select them i.e. \emph{exploit} such actions, to increase the reward value. But to know which actions are most rewarding, it has to \emph{explore} all possible actions many times. This is the trade-off between exploitation and exploration. 
The state space, which grows exponentially as the number of state variables increases, is another challenging aspect of RL. For the agent to learn effectively which action is
most rewarding in each state, it has to experience the same state multiple times, which is less likely as the state space grows. In such cases, the agent may
have to somehow generalize and use the experience gathered from similar states.
The third, known as the temporal credit assignment problem \cite{Sutton}, is that the agent should be able to reward or penalize past 
actions for each observed reward or penalty.

One way of addressing the temporal credit assignment problem is to maintain a Q-value of each state-action pair (s, a) under policy $\pi$ \cite{Sutton}.
It represents the expected value of the cumulative discounted future reward that is obtained when action \emph{a} is executed in state \emph{s}, and policy $\pi$ is followed subsequently.
One method to update Q-values is the SARSA update \cite{Sutton}.
\[ \delta = (r + \gamma Q(s_c, a_c) - Q(s_p, a_p)) \]
\[ Q(s_p, a_p) \leftarrow Q(s_p, a_p) + \alpha *  \delta \]
where $\delta$ is the Temporal Difference (TD) error.
In state $s_p$ when the agent takes action $a_p$, it receives a reward
\emph{r} and moves to state $s_c$. In state $s_c$ when it takes action $a_c$,
using the Q value of ($s_c$, $a_c$) the Q value of ($s_p$, $a_p$) is updated.
Here, $\alpha$ is the learning rate parameter and $\gamma$ is the discount rate parameter between 0 and 1.
In order to increase the cumulative sum of rewards, among the possible actions in state $s_c$, the action $a_c$ with the highest Q value of ($s_c$, $a_c$) is selected. Also, to explore the state-action space, with some probability, one of the possible actions is randomly selected.
The Q value of each state-action pair is initialized to $r_{max} / (1 - \gamma)$, where $r_{max}$ is the maximum (one-step) reward possible. This choice of 'optimistic' initialization enables the agent to explore each action sufficiently.

\subsection{Genetic Algorithm}
Genetic Algorithms (GA) \cite{GA} are search heuristics based on the ideas of natural selection and are used to generate near-optimal solutions to optimization problems. 
The search heuristic starts
with an initial population of candidate solutions. 
A solution is characterized by a set of properties (or chromosomes), where each property can take a range of values.
Typically solutions are represented using an array of bits where each bit corresponds to a property of the solution. This representation simplifies the crossover and mutation operations used 
for generating the next generation of solutions.
The fitness value of each solution is evaluated using an objective function obtained from the optimization problem being solved. A new generation of solutions is created from the current generation by combining solutions based on their fitness values. 

The crossover operation involves combining the properties of two parent solutions and creating two children solutions. 
Assume, P1 and P2 are two parent solutions, and C1 and C2 are the two children generated using the crossover operation. 
Child solution
C1 gets values from parent P1 for properties up to the crossover point and that of P2 after the crossover point, whereas, C2 gets values of P2 properties up to the crossover point and that of P1 after the crossover point. Each child solution
is then mutated i.e. one of its properties is modified with a small probability.

The process of creating new generation and evaluating them is repeated till either a satisfactory fitness level has been reached for the population or a maximum number of generations has been produced.

\section{Motivation}
\label{sec:Motivation}
\begin{figure}
\includegraphics[scale=0.32]{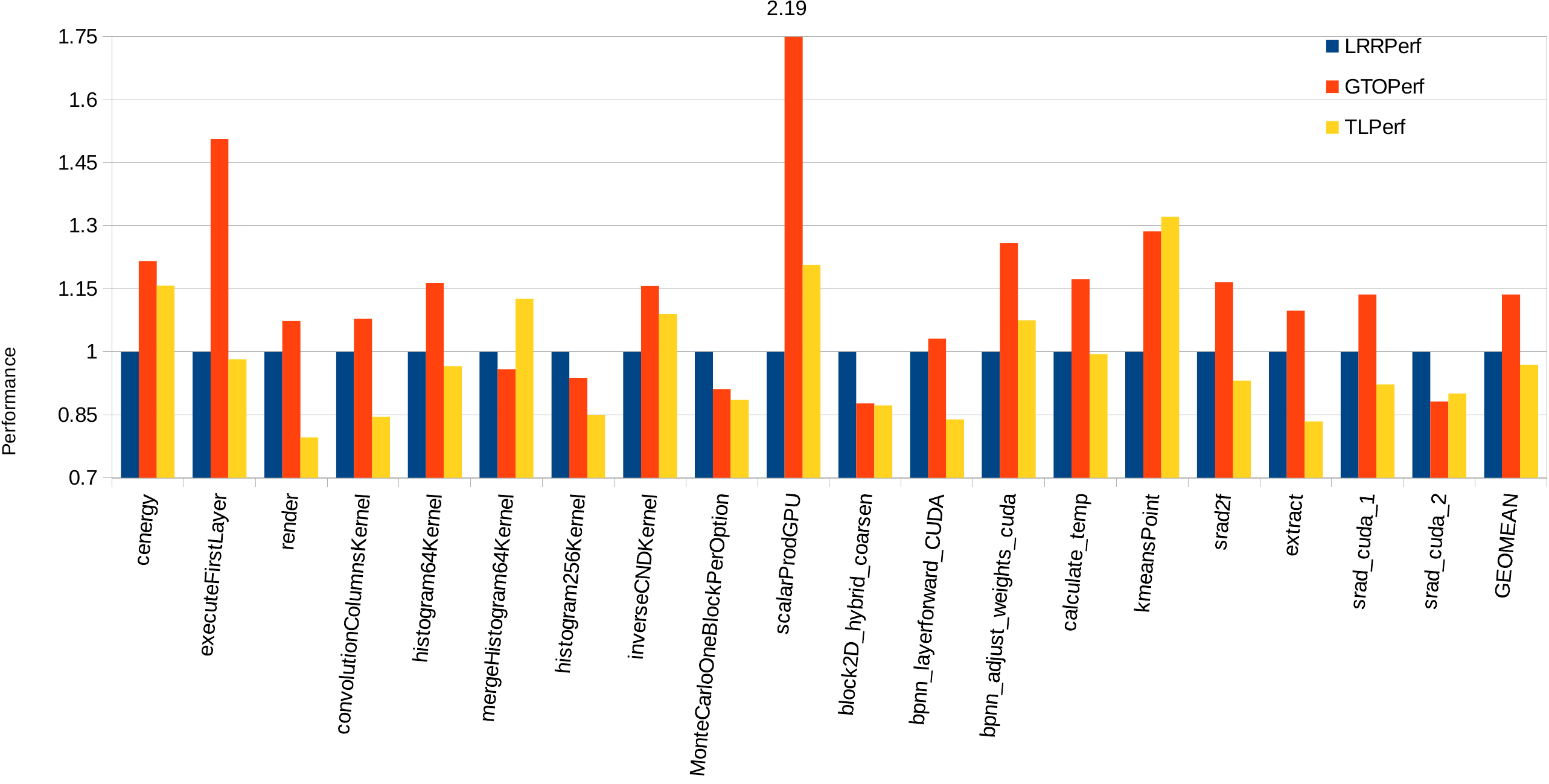}
\caption{Performance of GTO and TL normalized to LRR}
\label{motivation}
\end{figure}

Figure~\ref{motivation} shows data for the LRR, GTO and TL schedulers \cite{Gebhart} \cite{TwoLevel} from our simulation methodology explained in section \ref{sec:Exp}.
LRR gives equal priority to each warp and hence all warps on an SM tend to make equal progress. 
The advantage of LRR is that if warps of a TB access
nearby locations in memory, they will benefit from spatial locality while accessing L1-D/L2 caches and 
DRAM pages (row buffer hits). But the disadvantage is that, since all warps tend to make equal progress, they will
reach long latency instructions, such as global memory accesses, SFUs, etc., close to each other
in time.  As more and more warps issue long latency instructions close to each other in time, 
the warp scheduler will have fewer warps to hide the latency. 

As a solution to this problem, Narasiman et. al. \cite{TwoLevel} proposed the \emph{Two Level} warp scheduler which divides the warps in an SM into
\emph{fetch groups} and schedules the warps in each fetch group
in round robin manner. This can prevent the fetch groups from reaching long latency instructions close to each other in time. Figure ~\ref{motivation} shows the performance of GTO and TL normalized to LRR for a subset of benchmarks. 
TL performs the best in \emph{mergeHistogram64Kernel} and \emph{kmeansPoint} kernels. But as can be seen from Figure~\ref{motivation}, TL performs significantly worse for a number of kernels (e.g. \emph{render}, \emph{extract}).
One potential reason for
this is the fetch group size. One fixed fetch group size may not work
well for all workloads. 

GTO scheduler tries to prevent the problem due to long latency instructions by letting warps to make unequal progress. 
Since warps reach long latency instructions at different times, their latencies can be hidden in a better way.
As can be seen from the figure, even though the geometric mean performance of GTO is more than the other two schedulers, in kernels \emph{mergeHistogram64Kernel} and \emph{srad\_cuda\_2} its performance is the least among the three schedulers.
Both GTO and TL use heuristics to avoid stalls due to long latency 
instructions but still in kernels such as \emph{histogram256Kernel}, \emph{MonteCarloOneBlockPerOption}, \emph{block2D\_hybrid\_coarsen} and \emph{srad\_cuda\_2} LRR performs better than GTO and TL. 

In addition to long latency instructions, some other reasons for stalls are warp divergence \cite{WarpLevelDivergence}, allocation and deallocation of threads at TB granularity,
limited resources such as registers and shared memory available in SMs. 
As there are different reasons for stalls, it is extremely difficult to design 
one heuristic or combine multiple heuristics into one scheduling policy, that can schedule
warps in an intelligent way to reduce stalls in different types of workloads
and also different execution phases in a workload. 

As mentioned in the introduction, a special characteristic of GPU execution is that TBs of a kernel exhibit similar behavior (in terms of the dependency, control flow, synchronization, and stalls experienced across warps). Further the same kernel may be executed repeatedly in a loop. Hence it may be possible to learn some traits of workloads from the initial set of threads. 
We believe it is possible to design a scheduler that uses online learning to schedule warps in such a way that the number of stall cycles is reduced. The advantage of such a technique is that it can respond not only to different types of workloads but also to different execution phases in a workload.

\section{RLWS: Reinforcement Learning Based Warp Scheduler}
\label{sec:RLScheduler}
In this section we describe in detail our proposed RL based warp scheduler. 
\subsection{RL Agent}

In RLWS, the warp scheduler is the agent. Every cycle the agent has to select a warp to schedule. In Fermi GPU, each SM can have at most 48 resident warps and so each of the two warp schedulers can have at most 24 warps to select from \cite{Fermi}. 
As the number of actions increases, the size of state-action pair table also increases which causes the time needed for the agent to explore all possible state-action pairs, and exploit the most rewarding state-action pairs, to go up.

Table~\ref{table:pipeActionTable} shows the actions which identify the pipeline to schedule a ready warp to --- we refer to these as actions of type SELECT\_PIPELINE.
Fermi GPU contains 4 different types of memories viz., Global, Shared, Texture and Constant memory \cite{Fermi}. The average access latency of global memory is usually 1 to 2 orders of magnitude higher than Shared/Texture/Constant (STC) memory types. Hence two separate memory access actions were introduced in Table~\ref{table:pipeActionTable} to differentiate between them. 
Action NO\_INSTR does not issue any warp. Even though this action causes a stall cycle, it models being non-greedy, which may be beneficial sometimes. When at least one of the other actions is possible, action NO\_INSTR is not repeated in
consecutive cycles.
The idea behind selecting an action of type SELECT\_PIPELINE is to be able to learn dynamically which pipeline needs to be prioritized and issue warps to that pipeline.

\begin{table}
\scriptsize
\caption{Actions of type SELECT\_PIPELINE}
\begin{tabular}{c c}
\hline\hline
Action & Description \\
\hline
NO\_INSTR     &	Schedule no warp \\
SP\_INSTR     & Schedule a warp to SP pipeline \\
SFU\_INSTR    & Schedule a warp to SFU pipeline \\
GMEM\_INSTR   & Schedule a warp accessing global memory to MEM pipeline \\
STCMEM\_INSTR & Schedule a warp accessing STC memory to MEM pipeline\\
\hline
\end{tabular}
\label{table:pipeActionTable}
\end{table}

\subsection{RL Environment}
The environment of an RL based warp scheduler comprises of the warp queues, 
execution pipelines, caches, etc. Table~\ref{table:attrTable} shows the state attributes which we identified to represent the environment. Most of the attribute names are self explanatory. A state variable corresponding to ATBWB is true if there is at least one TB with some warps but not all, waiting at a barrier. 
If there are TBs which are yet to be assigned to an SM, then the variable corresponding to TBW is true. If threads of a warp diverge at a branch, then that
warp is considered to be a split(diverged) warp. NIW is used to count how many warps do not have any valid instruction in their instruction buffers. Variable corresponding to GNMIE counts number of issued but not completed memory instructions. Attribute RLMI is true if there is any warp with a memory instruction that has been observed to be a long latency memory instruction, whereas, attribute RGMI is true if there is any warp with a global memory access instruction. Attribute NRAI counts number of ready warps with SP and SFU instructions.

The last column shows
the range of values of each state variable. Some state variables can take a large number
of values, e.g. L1-D cache miss percentage can be any value between 0 and 100. 
In order to reduce the state space explosion, the value ranges of state variables are 
grouped into a small number of buckets (or sub-ranges) -- typically 2 to 4 -- and each  group is represented by a small integer value, e.g., from 0 to 3.
The range of values is not divided equally among the buckets. Either increasing or decreasing sizes of value ranges for each bucket was used depending on the state variable.  For example, if 3 buckets are used for L1-D miss percentage state variable, then 0-50\% is represented by bucket 0, 51-80\% by bucket 1 and the last 20\% by bucket 2. On the other hand, for the state variable NRGMI (Number of Ready Global Memory Instructions), increasing sizes of value ranges were used, e.g., with 4 buckets, 0-9\% was represented by bucket 0, 10-29\% by bucket 1, 30-59\% by bucket 2 and 60-100\% by bucket 3. 

\begin{table}
\scriptsize
\caption{Attributes to represent state of the environment}
\begin{tabular}{c c c}
\hline\hline
AttrName & Description & Value Range  \\
\hline
ATBWB   & Any TB with Warps at Barrier                   & 0-1 \\
ATBWF   & Any TB with Warps Finished                     & 0-1 \\
TBW     & TBs Waiting to be assigned to SMs              & 0-1 \\
RLMI    & Any warp with Ready Long Latency Memory Instr  & 0-1 \\
RGMI    & Any warp with Ready Global Memory Instr        & 0-1 \\
RSTCMI  & Any warp with Ready STC Memory Instr           & 0-1 \\
RSFI    & Any warp with Ready SFU Instr                  & 0-1 \\
RSPI    & Any warp with Ready SP Unit Instr              & 0-1 \\
NTF     & No TB Finished                                 & 0-1 \\
NIW     & Number of Idle Warps                           & 0-24 \\
NSW     & Number of Split(Diverged) Warps                & 0-24 \\
NFSFI   & Number of Warps with Next Instr as SFU Instr      & 0-24 \\
NFMI    & Number of Warps with Next Instr as a Memory Instr & 0-24 \\
NRSPI   & Number of Warps with Ready SP Instr               & 0-24 \\
NRGMI   & Number of Warps with Ready Global Memory Instr    & 0-24 \\
NRSTCMI & Number of Warps with Ready STC Memory Instr       & 0-24 \\
NRSFI   & Number of Warps with Ready SFU Instr              & 0-24 \\
NWI     & Number of Warps waiting as operands not ready     & 0-24 \\
NPS     & Number of Warps stalled as pipelines are full 	& 0-24 \\
NMPS    & Number of Warps stalled due to busy MEM Pipeline & 0-24 \\
NSFPS   & Number of Warps stalled due to busy SFU Pipeline & 0-24 \\
NSPPS   & Number of Warps stalled due to busy SP Pipeline  & 0-24 \\
NAIPMI  & Number of ALU Instrs Issued per Memory Instr 		& 0-24 \\
NRI     & Number of Warps with Ready Instr                  			& 0-24 \\
NWS     & Number of Schedulable Warps 									& 0-24 \\
NRAI    & Number of Warps with Ready ALU Instr              			& 0-24 \\
STBRMI  & Number of Warps with Ready Memory Instr With      			& 0-24\\
        & Same TB as Last Memory Instr Issued & \\
SMNMIE  & Number of Memory Instrs Executing on SM        & 0-40 \\
ICMP    & Instr Cache Miss Percentage              & 0-100 \\
L1MP    & L1-D cache Miss Percentage                       & 0-100 \\
L2MP    & L2 cache Miss Percentage                       & 0-100 \\
NIPL1M  & Number of Instrs Issued per L1-D Miss 	 & 0-100 \\
AGML    & Average Global Memory Latency					 & 0-800 \\
GNMIE   & Number of Memory Instrs Executing on GPU       & 0-600 \\
\hline
\end{tabular}
\label{table:attrTable}
\end{table}

\subsection{Reward and Penalty}
For the RL agent to learn which actions are beneficial, it needs to be rewarded or penalized for its actions. 
In our proposed warp scheduler, the RL agent is rewarded (reward value 1) for every cycle in which it is able to schedule a warp and penalized (reward value 0) for every stall cycle. 
\subsection{Q Value Table}
The Q value table maintains the expected value of the cumulative discounted future rewards. It is indexed using the state-action pair. So, if the environment is in state \emph{s} then the long-term value of taking action \emph{a}, is represented by Q(s,a).
If there are \emph{N} state variables, each can take $V_i$ distinct values and \emph{M} is the number of actions, then the size of the Q value table will be $V_1 * V_2 * \cdots * V_N *M$. For 7 state variables each with 4 buckets and 5 actions,
the Q value table size is 81920, requiring 327680 bytes of storage.

\subsection{Function Approximation}
\label{sec:FA}
Two problems with a Q value table are the size of the table and the time needed to populate the table. When the Q value table is very large, generalization 
from previously experienced states to ones that have not been experienced yet can be used. One technique to achieve generalization is Function Approximation, in which, Q(s, a) is approximated using a parametric function. The number of parameters is typically very small
compared to the number of state action pairs. We used a linear function of the parameters:
\[ Q(s, a) = \phi(s, a)^T\theta,\]
where, $\phi$ is a vector of features and $\theta$ is a vector of real valued weights \cite{FA}. 
The sizes of both $\phi$ and $\theta$ are equal to the number of state variables times the number of actions. For example, if there are 3 state variables \emph{s1, s2} and \emph{s3}, and two possible actions in each state \emph{a1} and \emph{a2}, then the state \emph{s} is represented by \emph{s = (s1, s2, s3)} and 
\[\phi(s, a1)^T = (f1(s1), f2(s2), f3(s3), 0, 0, 0), \] 
\[\phi(s, a2)^T = (0, 0, 0, f4(s1), f5(s2), f6(s3)), \] 
where functions $f1$ to $f6$ compute feature values. We used $1/(2^v)$ as the function for all features, where \emph{v} is the state variable value. Feature values as powers of 2 allows use of shifters instead of multipliers in computations of Q and $\theta$ values.
The weight vector is updated using the TD error, $\delta$ as shown below:
\[\theta = \theta + \alpha * \delta * \phi(s, a)\]
Since, in the above equation, feature values corresponding to only the chosen action can be non-zero, every time $\theta$ vector is updated, at most $N$ components of the vector will get updated, where $N$ is the number of state variables.

\subsection{Action-selection}
\label{sec:actionSelection}
The agent has to select a warp to schedule every cycle. It does so by selecting either the action with the highest Q value (exploitation) or a randomly chosen action (exploration), decided by the exploration rate. Higher Q value signifies more rewards and hence in order to increase rewards, the RL agent exploits actions with the highest Q values. 
In case of multiple warps with the same action, from among them, the agent selects the warp which was scheduled in the last cycle if it matches the chosen action, otherwise the oldest matching among them.

As explained before, the agent has to explore all possible actions to identify the most rewarding action in each state. This is achieved by selecting a random action instead of the action with the highest Q value, with a small probability. This probability is also known as the exploration rate \cite{Sutton}.

\subsection{Two Phases of Kernel Execution}
\label{sec:TwoPhases}
The execution of a kernel can be divided into two phases; first phase from the beginning of the kernel till there is at least one TB yet to be assigned to an SM and second phase from the time the last TB is assigned to an SM till completion of the kernel. 

RLWS treats these two phases differently.
The first phase has a fixed number of TBs running all the time and is usually longer than the second phase. In this phase RLWS gradually reduces both the exploration rate and the learning rate as time progresses.
In the second phase, the number of TBs and hence the number of warps keeps decreasing, hence, both the rates are held constant at their initial values.

\subsection{Search for RL parameters and attributes using GA}
As discussed before there are many factors such as the RL parameters, state variables, rewards and penalties, on which the performance of RLWS depends. By choosing a subset of state variables, a set of specific values for the RL parameters (alpha, gamma, exploration rate, etc.), a specific RLWS can be defined. Thus for the given set of state variables, RL parameter values and actions a family of RLWS can be defined. As the design space of RLWS is very large, instead of choosing a specific set of values for the RLWS in an ad hoc manner, we use GA to explore the design space systematically and choose the best performing RLWS.

The initial set of candidate solutions consists of 100 randomly generated solutions. 
The fitness of a solution is computed using the number of simulation cycles as reported by the GPGPU-SIM \cite{GPGPUSIM} simulator. For each candidate solution the fitness function computed the geometric mean speedup of the RLWS with the chosen subset of state variables and RL parameters, over the baseline scheduling algorithm on a set of benchmarks. 

The next generation solutions consisted of 90 solutions generated  using crossover and mutation operations on solutions in the current generation and 10 randomly generated solutions. To generate a pair of child solutions using the crossover and mutation operations, a pair of parent solutions (from the current generation) is selected. The probability of a solution getting selected is directly proportional to its fitness value. The pair of solutions generated using the crossover operation were mutated with a probability inversely proportional to the fitness values of their parent solutions. We used one crossover point and at most one mutation. In order to avoid the search getting stuck in a local minimum, we introduced 10 randomly generated solutions. Every 10th generation, instead of 10 random solutions, we reintroduced the best 10 solutions seen so far to keep properties of the best solutions present in the population. 

Each solution consisted of multiple state variables (Table~\ref{table:attrTable}) and RL parameters, such as, alpha, gamma, exploration rate, penalty and reward. We used GA not only to select which state variables, but also for their bucket sizes. We explored multiple values for each of the RL parameters. 

\subsection{Hardware Cost}

As mentioned before, since the Q table size can be very large, we used function
approximation to reduce the cost. For the following discussion, assume there are
\emph{N} state variables and \emph{A} possible actions in each state. \emph{N} registers are used to store values of the \emph{N} state variables. Since feature value corresponding to a state variable is obtained by raising 2 to the negative of the state variable value, we do not need to store the feature values explicitly. Feature values are either multiplied with $\delta$ or with $\theta$ vector, in which case the multiplication is replaced by shifting the operands with the corresponding state variable. For example, $(\theta1 * \phi1) \Rightarrow (\theta1 >> v1)$, where $v1$ is the value of the variable corresponding to $\phi1$.
To compute the temporal difference $\delta$, Q value of the previous state action pair is needed, which needs one register. At every time step, the vector of weights $\theta$ needs to be updated. Since the vector $\theta$ has \emph{N} * \emph{A} components, that many registers are needed to hold the $\theta$ vector. In addition to this, values of the 3 RL parameters, viz., learning rate, exploration rate and discount factor need to be stored. So, the total storage needed is $N + 1 + N * A + 3 = N * (A + 1) + 4$ registers per SM. We used the same $\phi$ and $\theta$ vectors for the two warp schedulers on each SM of the Fermi GPU. 
As discussed in the experimental section, as RLWS used 8 state variables and 5 actions, it needs 52 registers (208 bytes) as extra storage per SM.

Every cycle the warp scheduler either finds the action corresponding to the highest Q value from one among $A$ different actions, or chooses an action randomly, decided using the exploration rate parameter. To compute the Q value of a state action pair, $N$ multiplications and $N-1$ additions need to be performed. The $N$ multiplications are needed to multiply the $N$ relevant $\theta$ and $\phi$ vector components. As mentioned before, the values chosen for features are 
powers of 2 and hence we can use shift operations instead of multiplications.
From the 5 Q values corresponding to the 5 actions, RLWS finds the action with the highest Q value.
Also, every cycle $N$ components of $\theta$ need to be updated, which needs $N+1$ multiplications and $N$ additions. Since one of the operands of $N$ of these multiplication operations is the feature value, we can use shift operations.

\section{Experimental Evaluation}
\subsection{Experimental Methodology}
\label{sec:Exp}
We used GPGPU-Sim \cite{GPGPUSIM} simulator version 3.2.2 to evaluate our proposed RLWS. Table ~\ref{table:gpuconfig} shows the GPU configuration used in our simulation. 
We compiled all applications using NVCC version 4.2 with default optimization level and used the PTX code for
simulation.

We used benchmarks from GPGPU-SIM \cite{GPGPUSIM}, Parboil \cite{Parboil}, CUDA SDK \cite{SDK}, and Rodinia \cite{Rodinia} benchmark suites.
Table~\ref{table:bms} shows the list of kernels, number of TBs in the grid and number of resident TBs on the GPU. 
We used a large and diverse set of kernels to evaluate our proposed approach to make sure that the benchmarks cover varied characteristics of applications, such as memory and branch divergence, compute versus memory intensive, cache sensitive, irregular accesses, warp divergence, etc. We compare performance of RLWS with 4 schedulers viz., LRR, TL, GTO and iPAWS.

\begin{table}
\scriptsize
\caption{GPGPU-Sim Configuration}
\centering
\begin{tabular}{c c}
\hline\hline
Architecture & NVIDIA Fermi GTX480 \\
Number of SMs & 15 \\
Max Number of TBs per SM & 8 \\
Max Number of Threads per Core & 1536  \\
Shared Memory per Core & 48KB \\
L1-Cache per Core & 16KB \\
L2-Cache & 768KB \\
Max Number of Registers/Core & 32768  \\
Number of Warp Schedulers & 2 \\
DRAM Scheduler & FR-FCFS  \\
\hline
\end{tabular}
\label{table:gpuconfig}
\end{table}

\begin{table}
\scriptsize
\caption{Benchmarks}
\centering
\begin{tabular}{c c c c}
\hline\hline
Application & Kernel &  TBs & GPU Residency \\
\hline
AES & aesEncrypt128 & 257 & 90 \\
BFS & Kernel & 256 & 90 \\
CP & cenergy & 256 & 120 \\
LIB & Pathcalc\_Portfolio\_KernelGPU2 & 64 & 64 \\
LIB & Pathcalc\_Portfolio\_KernelGPU & 64 & 64 \\
LPS & GPU\_laplace3d & 100 & 100 \\
MUM & mummergpuKernel & 196 & 75 \\
NN & executeFirstLayer & 168 & 105 \\
NN & executeSecondLayer & 1400 & 120 \\
NN & executeThirdLayer & 2800 & 120 \\
NN & executeFourthLayer & 280 & 120 \\
NQU & solve\_nqueen\_cuda\_kernel & 256 & 45 \\
RAY & render & 512 & 75 \\
STO & sha1\_overlap & 384 & 75 \\
\hline
BlackScholes & BlackScholesGPU & 480 & 120 \\
convSeparable & convolutionRowsKernel & 18432 & 120 \\
convSeparable & convolutionColumnsKernel & 9216 & 120 \\
histogram & histogram64Kernel & 4370 & 120 \\
histogram & mergeHistogram64Kernel & 64 & 64 \\
histogram & histogram256Kernel & 240 & 120 \\
histogram & mergeHistogram256Kernel & 256 & 90 \\
MonteCarlo & inverseCNDKernel & 128 & 120 \\
MonteCarlo & MonteCarloOneBlockPerOption & 256 & 90 \\
scalarProd & scalarProdGPU & 128 & 90 \\
\hline
bfs & BFS\_in\_GPU\_kernel & 1 & 1 \\
bfs & BFS\_kernel\_multi\_blk\_inGPU & 14 & 14 \\
cutcp & cuda\_cutoff\_potential\_lattice6overlap & 121 & 120 \\
mri-q & ComputeQ\_GPU & 128 & 75 \\
sad & mb\_sad\_calc & 1584 & 120 \\
sad & larger\_sad\_calc & 99 & 99 \\
sgemm & mysgemm & 10 & 10 \\
stencil & block2D\_hybrid\_coarsen\_xff & 64 & 64 \\
tpacf & gen\_hists & 201 & 45 \\
\hline
backprop & bpnn\_layerforward\_CUDA & 4096 & 90 \\
backprop & bpnn\_adjust\_weights\_cuda & 4096 & 90 \\
bfs & Kernel & 1954 & 36 \\
bfs & Kernel2 & 1954 & 36 \\
b+tree & findRangeK & 6000 & 45 \\
b+tree & findK & 10000 & 45 \\
cfd & cuda\_initialize\_variables & 1212 & 120 \\
cfd & cuda\_compute\_step\_factor &1212 & 120 \\
cfd & cuda\_compute\_flux &1212 & 45 \\
cfd & cuda\_time\_step & 1212 & 120 \\
hotspot & calculate\_temp & 1849 & 60 \\
kmeans & invert\_mapping & 1936 & 90 \\
kmeans & kmeanPoint & 1936 & 90 \\
lavaMD & kernel\_gpu\_cuda & 1000 & 60 \\
lud & lud\_diagonal & 1 & 1 \\
lud & lud\_perimeter & 3 & 3 \\
pathfinder & dynproc\_kernel & 463 & 90 \\
srad\_v1 & srad & 450 & 45 \\
srad\_v1 & srad2 & 450 & 45 \\
srad\_v1 & reduce & 450 & 45 \\
srad\_v1 & extract & 450 & 45 \\
srad\_v1 & prepare & 450 & 45 \\
srad\_v1 & compress & 450 & 45 \\
srad\_v2 & srad\_cuda\_1 & 16384 & 90 \\
srad\_v2 & srad\_cuda\_2 & 16384 & 90 \\
\hline
\end{tabular}
\label{table:bms}
\end{table}

\subsection{Identifying RLWS Configuration}
We use GA to identify the RL parameters, attributes and number of buckets per attribute.
The top 9 attributes shown in Table~\ref{table:attrTable} are boolean which means they need two discrete values (buckets). The remaining 25 attributes can take a large number of values and we used 3 different bucket sizes viz., 2, 4 and 8 to represent them.

The GA based search mechanism also explored values for the 5 RL parameters viz., $\alpha$, exploration rate, $\gamma$, reward and penalty. For each of these we explored at least 5 different values. 
The number of possible solutions grows exponentially with the number of attributes and RL parameters --- for our setup the number of possible solutions is about $2^{70}$.

Evaluating the fitness of each candidate solution generated by GA needed simulating all of the kernels by representing the RL environment using the attributes and parameters identified by the solution. The fitness of a candidate solution is the geometric mean of performance of RLWS, using the configuration represented by the candidate solution. 
Of all the solutions found by GA, about 2/3rd were within 2\% range of the best solution.
Since the simulation of each kernel using GPGPU-Sim could take several hours, and several thousand candidate solutions need to be explored by the GA, in order to reduce the design exploration time, we used a subset of kernels (15 out of 59)\footnote {The subset of kernels was used only for exploring the RL design space. Performance results are reported for the entire set of 59 kernels.}. This subset, consisting of short running equivalent kernels for the long running ones, was chosen after careful experimentation, identifying kernels which have similar behaviour for different values of RL parameters and attributes.

From all the solutions generated by GA in 100 generations (nearly 10,000 solutions), we selected the top 10 solutions and computed their fitness using all 59 kernels. Table~\ref{table:rlParamEnv} shows the values of RL parameters and state variables of the best solution among these 10 solutions. The performance numbers reported in Figure~\ref{perf} were obtained using these RL parameter and attribute values.
Note, the overhead of running GA and identifying the RLWS parameters is only once per GPU configuration.

The second part of table~\ref{table:rlParamEnv} shows the attributes and number of buckets as selected by the GA to represent the environment. Every cycle RLWS computes values of these attributes. 
The second column of this table shows the number of buckets used for each attribute. Attributes AGML, GNMIE and L2MP are common for all SMs on a GPU, whereas, the others are SM specific. The storage overhead of each attribute is at most a couple of 4 byte wide registers.

\begin{table}
\scriptsize
\caption{RL Parameter and State Variable values}
\centering
\begin{tabular}{c c}
\hline\hline
Name & Value\\
\hline
Learning Rate & 0.09 \\
Exploration Rate & 0.04 \\
Discount Factor & 0.95 \\
Penalty & 0 \\
Reward & 1 \\
\hline
Average Global Memory Latency (AGML) & 2 \\
Number of Memory Instrs Executing on GPU (GNMIE) & 8 \\
L1-D Miss Percent (L1MP) & 8 \\
L2 Miss Percent (L2MP) & 2 \\
Number of Warps with Next Instr as a Memory Instr (NFMI) & 4 \\
Number of Instructions Issued per L1-D Miss (NIPL1M) & 4 \\
Number of Warps with Ready ALU Instr (NRAI) & 4 \\
Number of Memory Instrs Executing on an SM (SMNMIE) & 4 \\
\hline
\end{tabular}
\label{table:rlParamEnv}
\end{table}

\subsection{Performance Analysis}
\begin{figure}
\subfloat[Set 1] {\includegraphics[scale=0.30]{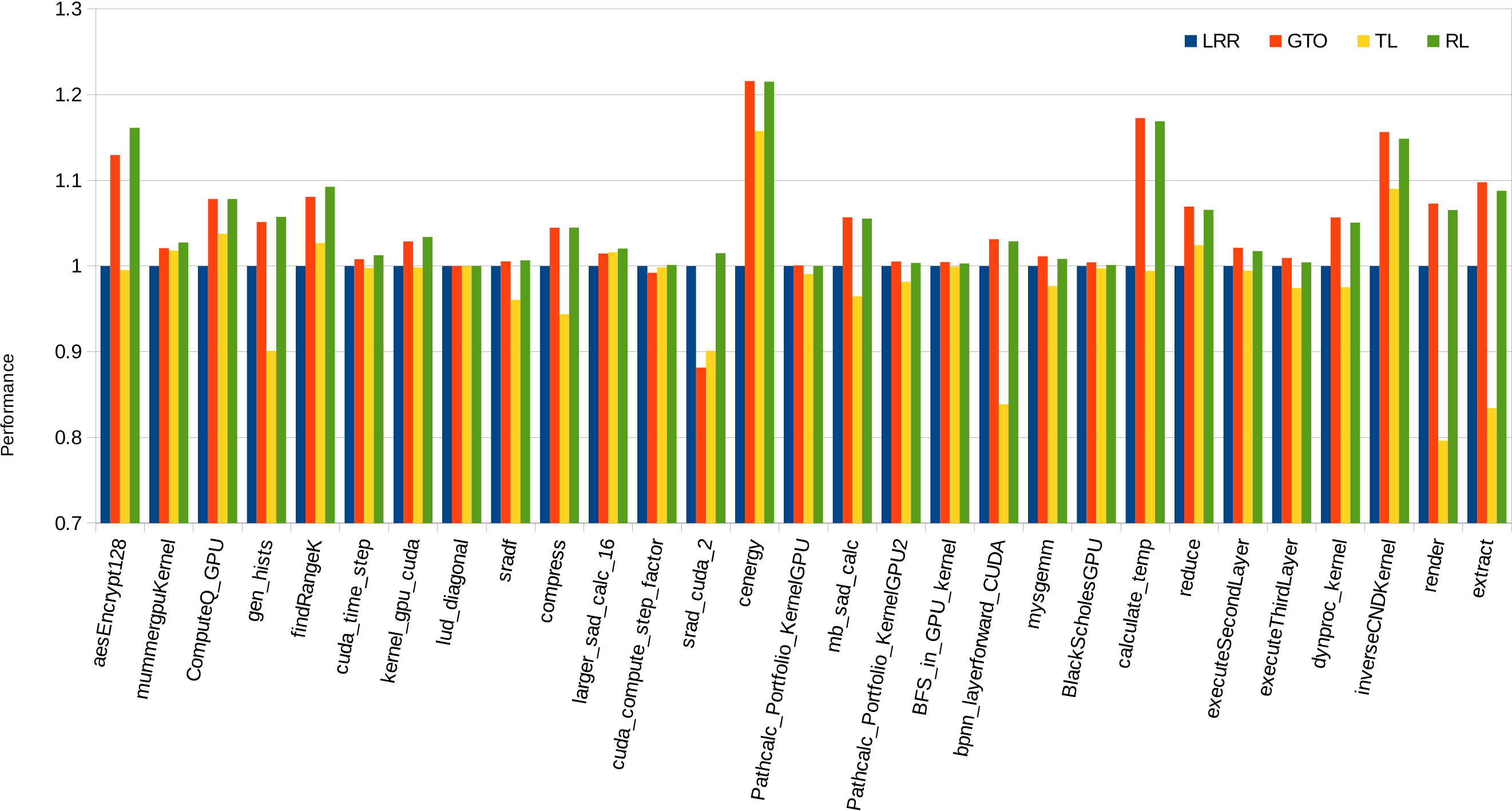}}
\newline
\subfloat[Set 2] {\includegraphics[scale=0.30]{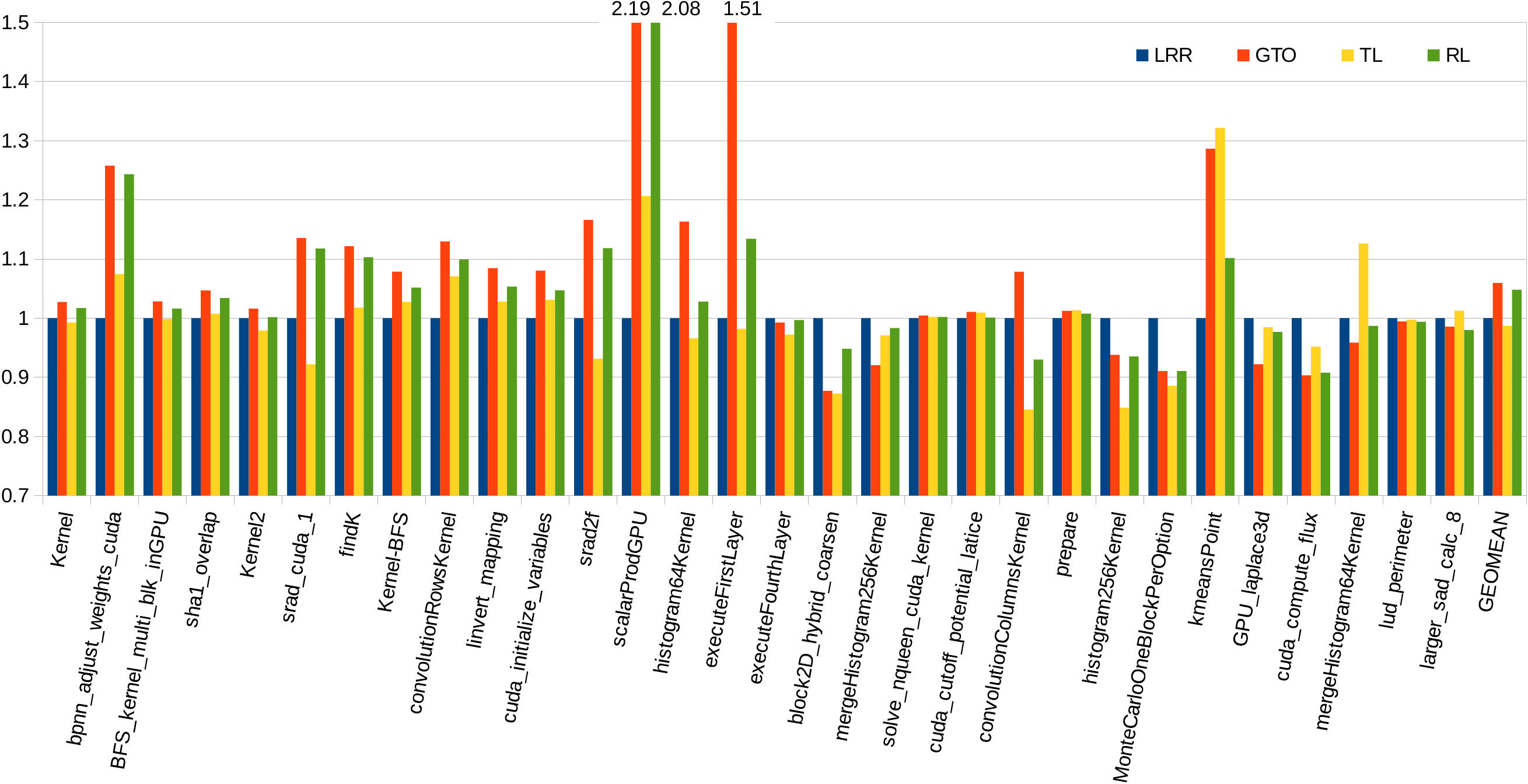}}
\caption{Performance of RLWS over LRR, TL and GTO}
\label{perf}
\end{figure}

Figures~\ref{perf}(a) and~\ref{perf}(b) show the performance results of RLWS over the three baseline warp schedulers viz., LRR, TL and GTO. We divided the kernels into two sets to make the graphs more readable. The column GEOMEAN in Figure~\ref{perf}(b) shows the geometric mean performance of RLWS over the three baseline scheduling algorithms. 
The performance improvement over LRR and TL are 1.05x and 1.06x respectively. It performs nearly as well as GTO (0.99x), which is due to the slowdown of more than 5\% in 4 out of the 59 kernels.

\begin{table}
\scriptsize
\caption{RLWS Rank and Performance}
\centering
\begin{tabular}{c c c c c}
\hline\hline
RLWS Rank & RLWS\_LRR & RLWS\_GTO & RLWS\_TL & Num Kernels\\
\hline
1 & 1.04 & 1.02 & 1.06 & 13 \\
2 & 1.08 & 0.98 & 1.09 & 34 \\
3 & 0.97 & 0.98 & 0.98 & 10 \\
4 & 0.99 & 1.00 & 0.98 & 2 \\
\hline
\end{tabular}
\label{table:rlRank}
\end{table}

Table~\ref{table:rlRank} shows the geometric mean performance of RLWS over the three baseline scheduling algorithms, sorted by the rank of RLWS. For each rank, the last column shows the number of kernels for which RLWS is at that rank (position in decreasing order of performance of all four schedulers). Columns 2, 3, and 4 show performance of RLWS w.r.t. LRR, GTO and TL respectively. So, the 2nd row shows that RLWS is at rank 2 (second best) for 34 kernels and its geometric mean performance improvement over TL on those 34 kernels is 1.09x. 

Figure~\ref{perf} and Table~\ref{table:rlRank} show that RLWS is a robust warp scheduler, in the sense, its performance is always close to the best. 
For about 80\% (47 / 59) kernels, RLWS is either the best or close to the best.
RLWS successfully figured out more rewarding actions for such a diverse set of kernels.

We compared the performance of RLWS with iPAWS \cite{iPAWS}, a more recent scheduler, which uses a heuristic based on the instruction issue pattern of warps to decide whether LRR or GTO is better for the kernel. Since our implementation of iPAWS based on the details in the paper \cite{iPAWS}, chose the correct scheduling policy for only 17 out of 59 kernels, we used an oracle analysis (iPAWS\_Oracle) also to choose between LRR and GTO.

Simulation data shows that, for 14 kernels LRR is better than GTO (LRR friendly) and for the remaining 45 kernels GTO is better than LRR (GTO friendly). Our implementation of iPAWS correctly chose LRR for the 14 LRR friendly kernels, but it chose GTO for only 3 out of the 45 GTO friendly kernels.
Table~\ref{table:rlIPAWS} shows, on LRR friendly kernels, RLWS, iPAWS and iPAWS\_Oracle show equal performance improvement of 3\% over GTO. The slowdown in iPAWS\_Oracle for LRR friendly kernels is due to the adapt and recover phases which do not use LRR scheduling policy.
On GTO friendly kernels, 
iPAWS shows a slowdown of 5\% due to wrong analysis and wrong choice of scheduling policy after the adapt phase. RLWS is better than LRR by 7\% and iPAWS by close to 2\%.
Overall, RLWS is better than iPAWS by close to 2\%. It is evident from this data that RLWS is able to learn a better scheduling policy on the LRR friendly kernels and outperform GTO.
\begin{table}
\scriptsize
\caption{Comparison of RLWS with iPAWS}
\centering
\begin{tabular}{c c c c c c}
\hline\hline
Kernel Type & LRR & GTO & iPAWS & iPAWS\_Oracle & RLWS\\
\hline
LRR friendly Kernels & 1 & 0.94 & 0.97 & 0.97 & 0.97\\
GTO friendly Kernels & 1 & 1.1  & 1.05 & 1.1  & 1.07 \\
All kernels          & 1 & 1.06 & 1.03 & 1.07 & 1.05 \\
\hline
\end{tabular}
\label{table:rlIPAWS}
\end{table}

\subsection{Meta-scheduling Action}
Next, we tried an alternative RLWS method, referred to as RLWS\_MS, which chooses meta-actions, rather than operations performed by a warp. 
RLWS\_MS  selects one of the actions shown in table~\ref{table:metaSchedActions} every cycle to identify the warp to be scheduled. 
We used GA to identify the right set of state variables and RL parameters that result in the best performing RLWS\_MS. The identified parameters, shown in table~\ref{table:rlParamEnvMS}, are different 
from RLWS.
The geometric mean performance of RLWS\_MS is 1.06x over LRR, 1.07x over TL and matches GTO. Table ~\ref{table:rlwtRank} shows that RLWS\_MS is the best or second best for about 80\% kernels. 

Tables~\ref{table:rlRank} and~\ref{table:rlwtRank} show that our proposed design is able to learn different types of actions on a diverse set of kernels. 

\begin{table}
\scriptsize
\caption{RL Parameter and State Variable values for Meta-scheduling Action}
\centering
\begin{tabular}{c c}
\hline\hline
Name & Value\\
\hline
Learning Rate & 0.01 \\
Exploration Rate & 0.01 \\
Discount Factor & 0.999 \\
Penalty & 0 \\
Reward & 1 \\
\hline
Any Thread Block with Warps at Barrier(ATBWB) & 2 \\
Any Thread Block with Warps Finished(ATBWF) & 2 \\
Number of ALU Instructions Issued per Memory Instruction(NAIPMI) & 8 \\
Number of Ready Global Memory Instructions(NRGMI) & 8 \\
Number of Split(Diverged) Warps(NSW) & 4 \\
Number of Schedulable Warps(NWS) & 2 \\
Any Warp with Ready SP Unit Instruction(RSPI) & 2 \\
\hline
\end{tabular}
\label{table:rlParamEnvMS}
\end{table}

\begin{table}
\scriptsize
\caption{Meta-scheduling Actions}
\centering
\begin{tabular}{c c}
\hline\hline
Action & Description \\
\hline
GTO & Schedule warps in GTO order \\
Youngest & Schedule the youngest ready warp \\
LRR & Schedule the next (round robin) ready warp  \\
Youngest Barrier & Schedule the youngest of warps with siblings waiting at barrier\\
Youngest Finish & Schedule the youngest of warps with siblings finished execution\\
\hline
\end{tabular}
\label{table:metaSchedActions}
\end{table}

\begin{table}
\scriptsize
\caption{RLWS\_MS Rank and Performance}
\centering
\begin{tabular}{c c c c c}
\hline\hline
Rank & RLWS\_MS\_LRR & RLWS\_MS\_GTO & RLWS\_MS\_TL & Kernels\\
\hline
1 & 1.08 & 1.01 & 1.10 & 28 \\
2 & 1.08 & 0.99 & 1.09 & 19 \\
3 & 0.96 & 0.98 & 0.98 & 12 \\
\hline
Overall & 1.06 & 1.00 & 1.07 & \\
\hline
\end{tabular}
\label{table:rlwtRank}
\end{table}

\subsection{Discussion}
In addition to RLWS and RLWS\_MS, we also tried a few other RLWS methods. 
We describe a few of them briefly here.
We designed a scheduler consisting of
two RL agents, one to select the pipeline to schedule to and the other to select a TB, and the final action was a combination of the two, i.e. a warp with instruction matching
the pipeline chosen by the first agent and from the TB as selected by the second agent. The performance with two RL agents was a little lower than the performance with a single RL agent. In addition to this, independently, we increased the set of actions in RLWS by including bypass L1-D cache, limiting the number of warps considered for scheduling, and combinations of these with the other actions using multiple RL agents, but they also had a little lower performance.

RLWS learns afresh for each kernel. 
We conducted a set of experiments in which the agent learns continuously across multiple kernels, iterated multiple times. In particular, 
the values of $\theta$ were remembered across these executions and updated. The results were
similar to the ones with learning afresh.
As a sensitivity study we reduced the frequency of choosing an action every 2, 4, 8 and 16 cycles with the same state variables and RL parameters as shown in Table~\ref{table:attrTable}. It showed performance degradation of less than 1\%. 

Even though RL based solutions have been successful in various fields and RLWS achieved reasonable performance, it could not outperform GTO. One reason is the precision of action chosen, i.e., how precisely it can identify a warp. For example, if the action is to choose a warp with current instruction as a memory instruction, it is not clear which one when there are multiple such warps. Increasing number of actions to improve precision results in a substantial increase in state-action space and hence increase in the cost and time to learn. 
Identifying the right set of actions to improve precision without increasing the number of actions will possibly improve performance of RLWS.

We believe, our work is a good initial study of using RL to solve the challenging problem of scheduling warps.

\section{Related Work}

Lee et al. proposed iPAWS \cite{iPAWS} which dynamically selects between a greedy and a round-robin warp scheduler, based on the instruction issue pattern. 
Once a scheduling policy is selected, it will be used till the end of the kernel. The learning phase can be a significant portion of execution for kernels invoked with a small number of TBs. In contrast, RLWS can have more actions and it can adapt to different execution phases of a kernel. RLWS is continuously learning and taking actions based on whatever it has learnt so far and hence even kernels with a small number of TBs, can benefit.
Phase Aware Warp Scheduler \cite{PhaseAware}, identifies phases of execution at compile time and embeds phase lengths in the instructions. 
The warp scheduler chooses the warp with the shortest length for its next phase.

A number of heuristic warp scheduling methods have been proposed \cite{Jog}, \cite{OWL}, \cite{Gebhart}, \cite{NMNL}, \cite{DAWS}, \cite{CCWS}, \cite{TwoLevel}, \cite{Mascar}, \cite{WarpedPreexecution}, \cite{PriorityCache}, 
each one attempting to address a specific factor, such as varying dynamically the thread level parallelism, reducing cache and memory contention, prioritizing critical warps, 
prioritizing memory requests from a set of warps, etc. These methods typically perform only for a subset of applications.

Yang et al. \cite{ShMemMult} and Tarjan et al. \cite{ondemand} discuss techniques to improve shared memory and register utilization respectively.
Xiang et al. \cite{WarpLevelDivergence} discuss allocation and deallocation of registers at warp level. CAWS \cite{CAWS} and CAWA \cite{CAWA} focus on prioritizing critical warps to reduce warp divergence.
Various hardware and software techniques to handle thread divergence have been proposed in \cite{ThreadFrontier}, \cite{DynWarpSubdivision}, \cite{DualPath}, \cite{PATS}, \cite{ThreadBlockCompaction}, \cite{DynWarpForm}, \cite{CAPRI}, \cite{SIMDlanePermutation}, \cite{VariableSizeWarp}, \cite{CCC}. 

Ipek et al. \cite{Ipek}
proposed a  RL based memory controller, which learns
to schedule DRAM commands based on a small number of system variables.
McGovern et al. \cite{McGovern} used reinforcement learning to schedule basic
blocks. 
\emph{Deep} neural networks have emerged as very successful predictive models for tasks such as object recognition and natural language understanding. 
Various proposals \cite{EIE}, \cite{Eyeriss}, \cite{vDNN}, \cite{DNNWEAVER}, discuss solutions to accelerate, minimize energy, and speed up the training process of neural networks.
Although in principle, we can use neural networks as a function
approximator within RLWS (in place of our linear scheme), they are not
ideally suited for being trained on-line.

\section{Conclusion}
We proposed a RL based warp scheduler. We compared performance of RLWS on a large set of 59 kernels against 4 warp schedulers viz., LRR, TL, GTO and iPAWS.
By adapting on-line to the characteristics of each specific workload, RLWS is found to work well "across the board", achieving either the best or close to the best performance on 80\% of kernels, proving its robustness. 
RLWS opens up new opportunities in designing warp schedulers. We leave exploring RLWS for scheduling concurrent kernels for future work.  Also, we plan to use advanced RL algorithms to compute Q values, identify more attributes to improve the environment representation, etc.

\end{document}